\documentclass[manuscript,nonacm]{acmart}
\AtBeginDocument{%
  }

\setcopyright{acmlicensed}
\copyrightyear{2025}
\acmYear{2025}
\acmConference[Presented at the Human-centered Explainable AI Workshop (HCXAI) @ CHI 2025, DOI: \href{https://doi.org/10.5281/zenodo.15170444}{10.5281/zenodo.15170444}]{April 26--May 1,
  2025}{Yokohama, Japan}

\begin{document}

\title{Legally-Informed Explainable AI}


\author{Gennie Mansi}
\email{gennie.mansi@gatech.edu}
\orcid{0000-0001-6186-5102}
\author{Naveena Karusala}
\email{nkarusala3@gatech.edu}
\author{Mark Riedl}
\email{riedl@cc.gatech.edu}
\affiliation{%
 \institution{Georgia Institute of Technology}
 \streetaddress{North Avenue}
 \city{Atlanta}
 \state{Georgia}
 \country{USA}
 \postcode{30332}
}

\renewcommand{\shortauthors}{Mansi, Karusala, and Riedl}

\begin{abstract}

Explanations for artificial intelligence (AI) systems are intended to support the people who are impacted by AI systems in high-stakes decision-making environments, such as doctors, patients, teachers, students, housing applicants, and many others. 
To protect people and support the responsible development of AI, explanations need to be \textit{actionable}---helping people take pragmatic action in response to an AI system---and \textit{contestable}---enabling people to push back against an AI system and its determinations. 
For many high-stakes domains, such as healthcare, education, and finance, the sociotechnical environment includes significant legal implications that impact how people use AI explanations. 
For example, physicians who use AI decision support systems may need information on how accepting or rejecting an AI determination will protect them from lawsuits or help them advocate for their patients.
In this paper, we make the case for Legally-Informed Explainable AI, responding to the need to integrate and design for legal considerations when creating AI explanations. We describe three stakeholder groups with different informational and actionability needs, and provide practical recommendations to tackle design challenges around the design of explainable AI systems that incorporate legal considerations.

\end{abstract}

\ccsdesc[500]{Human-centered computing~HCI theory, concepts and models}
\keywords{Legally-Informed Explainable AI, actionability, contestability, AI explanations, legal considerations, design challenges \\ \\ Presented at the Human-centered Explainable AI Workshop (HCXAI) @ CHI 2025, DOI: \href{https://doi.org/10.5281/zenodo.15170444}{10.5281/zenodo.15170444}}


\received{20 February 2025}

\maketitle

\section{Introduction}
Artificial intelligence (AI) systems are increasingly involved in high-stakes decision-making by humans, such as healthcare, housing, financial, and educational determinations~\cite{Lipton2018, ChariEtAl2020, GuidottiEtAl2018, DohnEtAl2020}. Even initial deployments of these systems have shown how error prone they are, from incorrectly prescribed treatments \cite{AIAAIC_Cancer} to incorrectly denying insurance coverage \cite{Statnews_Medicare} or reviewing applications \cite{AIAAIC_WorkdayJobs}. As a part of responsible AI development, many turn to  {\em Explainable AI} (XAI)---AI systems that provide human-understandable explanations for their reasoning and output~\cite{Lipton2018, ChariEtAl2020, GuidottiEtAl2018}. Explanations are intended to support the people---doctors, patients, teachers, students, housing applicants, and many others---who are impacted by decisions made by AI systems. 

Users, AI decisions, and the explanations generated by AI systems do not exist in a vacuum but are embedded in a complex sociotechnical environment \cite{PoquetAndDeLaat2021, Sabanovic2010, JonesEtAl2013}. Human-Centered Explainable AI~\cite{Ehsan2020HumancenteredEA} has focused on understanding people's needs to inform and improve explanations for an AI system's behavior. 
For example, past HCXAI Workshop papers have highlighted nuances around how explanations need to be tailored to specific users' context, such as those of older adults aging in place \cite{MathurEtAl2024_HCXAI} or of people performing manual tasks alongside an AI system in manufacturing \cite{WatkinsEtAl2024_HCXAI}. 

As the concept of explainability has evolved, researchers have emphasized that to achieve transparency and accountability, AI systems do not just need to communicate why or how a decision was made, they also need to enable pragmatic action~\cite{JornoAndGynther2018, LyuEtAl2016, ChoEtAl2019, JoshiEtAl2019, TanAndChan2016, WiratungaEtAl2021, SinghEtAl2021}. That is: they need to be {\em actionable}. An important class of actionability is {\em contestation}---a person's ability to push back against an AI system's use and its determinations. End user explanations can help individuals make better-informed decisions while calibrating their trust of the AI system \cite{GuidottiEtAl2018, MothilalEtAl2020, FerrarioEtAl2020, Spiegelhalter2020, GliksonAndWoolley2020} and expanding their range of actions~\cite{Mansi2023DoSomething}.

\textbf{AI systems, by virtue of making decisions more opaque, can tip the balance of power away from those who are more vulnerable and those advocating for the vulnerable.}
For example, in healthcare, physicians who use AI decision support systems may have the information at hand to push back against an AI system's recommendation to advocate for a patient or protect themselves from a malpractice lawsuit~\cite{Mansi2024_LawyersContestability}.
Housing applicants may not know an algorithm has violated their rights and experience increasing housing instability as a result~\cite{Karusala2024Contestability}.
Should perceived harms occur in the presence of AI decision-support, legal representatives may not be able to audit black box systems \cite{JinAndSalehi2024,buiten2023vision}. 
Further complicating matters, 
those who are affected by AI decisions are especially vulnerable when the laws pertaining to the AI decisions are prone to change suddenly and significantly or when the laws may not be responsive to those technological innovations~\cite{akpuokwe2024legal}. All of these scenarios point to the obfuscation of legal rights and options currently perpetuated by AI.

We argue for \textbf{Legally-Informed XAI}, responding to the need to design AI explanations to account for the legal concerns of those impacted by AI, to ultimately ensure actionability and contestability. 
An important theme in HCI research is the {\em power dynamics} that shape human-AI interactions. The deployment of AI decision-support tools increasingly shifts power toward those deploying the systems. People interacting with AI decision-support tools have reduced agency as a result but are often still responsible, accountable, or liable~\cite{Sundholm2024}.
The ability to contest AI decisions restores agency to users.
However, opaque systems are increasingly hard to contest because it is more challenging to get information demonstrating a system is making a mistake or breaking the law \cite{Statnews_Medicare}.
As a first step,
we highlight three groups of stakeholders---{\em decision makers}, {\em decision subjects}, and {\em legal representatives}---along with their legal considerations that need to be accounted for
to mitigate the power dynamics exacerbated by algorithmic decision-making tools.
We then provide practical recommendations to incorporate legal considerations into AI explanations for responsible AI development.

\section{Stakeholders, Actionability, and Contestability}
In this section, we detail the legal considerations of three groups of stakeholders from an XAI perspective.

\textbf{Decision makers} are people that act alongside AI systems to make decisions, serving as buffers between AI recommendations and more vulnerable populations. Some decision makers have a legal obligation to work in the best interests of others. For example, medical physicians are called on to provide oversight of AI medical recommendation systems, protecting patients from harmful AI decisions \cite{NyrupAndRobinson2022, AbramoffEtAl2020, FroomkinEtAl2019}. Decision makers could also include those with relatively weaker legal obligations to act in the best interests of another person, such as those processing job, loan, or housing applications.   
If decision makers are to protect against or avoid unwanted outcomes, then it is important to support decision makers in understanding their legal liability and contesting improper AI decisions or applications. 

\textbf{Decision subjects} are people directly subject to decisions made by an AI system or someone using AI decision-support. 
Users, developers, and deployers of AI systems sometimes use the black-box nature of the system to evade liability, leaving decision subjects without recourse~\cite{Statnews_Medicare,metcalf}. 
For decision subjects, contesting AI systems is fraught with power dynamics, particularly in contexts where decision makers are not incentivized to advocate for decision subjects. 
Examples of such contexts include tenant screening, loans, insurance claims processing, or public surveillance. 
Prior work~\cite{kapania2022because,ramesh2022platform} has found that culture and economic precarity can shape perceptions of AI, noting that mistrust of human institutions, feelings of obligation, or fear may result in undue trust in, or acceptance of, AI decisions. 
We assert that AI systems must provide sufficient information to enable contestation, and that this may often involve information about legal rights.

\textbf{Legal representatives} are people who represent decision subjects, decision makers, technology creators, and others seeking recourse through the courts for harms involving AI systems.
While the harmed individual faces a power imbalance in favor of the AI system's developers and deployers, the legal representative of the harmed party also faces a power differential: they may not have the ability to audit an AI system or to make sense of the raw audit data. For example, Jin and Salehi~\cite{JinAndSalehi2024} investigate public defenders scrutinizing computational forensic software and find barriers to building cases against them despite access to performance evaluations, such as difficulties understanding how these technologies are developed and used. 
Legal representatives can also benefit from explanations, but their informational needs differ from those of their clients.
Consequently, there are growing calls to recognize and involve legal representatives when understanding and designing AI systems.

\section{An Evolving Technological and Legal Landscape: A Healthcare Case Study}
In this section, we take a first step in drawing out the complexities of creating Legally-Informed XAI through an extended case study of medical AI decision systems.
Physicians, their patients, and legal representatives are caught in an evolving landscape of technology and law.  While we focus on medical AI, the principles we discuss are present in other domains as well, including education, finance, criminal justice, and privacy, among others \cite{AlfrinkEtAl2023_Framework, Karusala2024Contestability}. 

There are hopes that AI can improve many aspects of medical care, including workflows, diagnoses, and treatments \cite{GerkeEtAl2020}. However, the integration of AI in actual medical settings has had mixed success. Prominent concerns have been documented including: alert fatigue~\cite{WongEtAl2021}; incorrectly prescribing cancer treatments~\cite{AIAAIC_Cancer}; misdiagnosing heart attacks~\cite{AIAAIC_HeartAttacks, TechCrunch_HeartAttacks}; wrongly denying opioids~\cite{Wired_Drugs, AIAAIC_Drugs}; difficulty understanding AI recommendations~\cite{ClassenEtAl2018}; and
accentuation of historical and systemic bias for marginalized groups~\cite{Benjamin2019, AbramoffEtAl2020}. 

Policy changes from national to local levels also impact medical decision making. The last several years have seen large-scale policy changes in the United States in response to new technology, unprecedented medical events, and court rulings. A common failure model of machine learning systems in particular is the inability to respond to rapid changes to the operating environment that causes it to differ from the historical training data on which the system was trained~\cite{EE_EhsanEtAl2021}. In the medical domain, for example, the 2020 pandemic saw the emergency authorization of vaccines and treatments \cite{FDA_EmergencyAuthorization2020}. In 2021, the United States Supreme Court issued rulings on women's health, creating an uneven map of where medical procedures could be performed within the country \cite{DobbsVJackson_ConstitutionCenter}. Laws and regulations regarding FDA approval processes for medical devices and software incorporating AI also changed \cite{FDA_AIMedical}. Even hospitals' guidelines about treatment vary, such as the kinds of antibiotics or procedures that should be followed. Additionally, there are also regular changes around approved or recommended medications, medical devices, and procedures.

\subsection{Physicians' Legal Considerations}

Physicians' legal considerations are closely tied to their needs around \textit{actionable} and \textit{contestable} AI systems. Medical doctors are tasked with making decisions at the intersection of medical knowledge and patients' values and individual needs \cite{MolemanEtAl2021}. Physicians are and feel accountable for the well-being of their patients and their own livelihoods as they use medical AI systems. Physicians must often make nuanced decisions, such as helping patients decide between cancer treatments to support their quality of life or administering a medication for an unapproved use when other treatments are not working, which may not be recommended by an AI system. Physicians must also advocate for patients, sometimes going against AI recommendations that may be incorrect, or against insurance companies that unjustly deny coverage to patients \cite{Statnews_Medicare}. Access to legally salient information could help empower physicians to push back against opaque denials. However, a lack of design around, and access to, these details prevents physicians and others from successfully contesting decisions, leaving patients no choice but to sue for coverage \cite{Statnews_Medicare}.

Additionally, physicians need to protect themselves from legal liability if a patient is harmed by a decision with an AI system \cite{vcartolovni2022ethical,jones2023artificial}. Physicians' concerns around malpractice liability are pervasive and materially impact medical practices \cite{NashEtAl2004, VelthovenAndWijck2012}---physicians may order more tests and procedures, take more time to explain risks, restrict the scope of their practice, and refer patients to other specialists \cite{CarrierEtAl2010, Dickens1991, NashEtAl2004}. These practices do not seem to improve the quality of care \cite{VelthovenAndWijck2012} despite increased costs for patients \cite{CarrierEtAl2010, VelthovenAndWijck2012}. Further, both the threat of and actual legal claims are associated with negative psychological, physical, and behavioral impacts for physicians
\cite{NashEtAl2004}. 

\subsection{Patients' Legal Considerations}
As decision subjects, patients are vulnerable to the use of AI systems by both physicians and others such as insurance companies. XAI systems are typically tightly coupled with the AI black-boxes they are explaining. 
This means that explanations are likely geared towards the decision maker (landlords or property managers in this case). Meanwhile, the explanations given directly to decision subjects may not convey actionable information that would be to the benefit of the decision subject and to the disadvantage of the decision maker; explanations in this case may even be {\em dark patterns}~\cite{chromik2019darkXAI} or {\em explainability pitfalls}~\cite{Ehsan2021Pitfalls}. 

This risk is heightened because while there are incentives for companies to create AI explanations and systems that are actionable for decision makers such as physicians, they are not incentivized to do so for decision subjects. Insurance companies are already using AI systems to increase profits by using opaque algorithms to deny medically necessary treatments for vulnerable patients groups, such as the elderly or people with disabilities \cite{Statnews_Medicare}. Outside of the tangible physical harms patients face with these denials, they also bear many intangible emotional and psychological harms as they work to push back against unjust decisions. Consequently, decision subjects such as patients are more at risk of absorbing negative impacts with fewer paths for recourse outside of legal action.

\subsection{Legal Representatives' Considerations}
As AI systems are increasingly deployed in real world medical environments, there will invariably be undesired and unanticipated effects on decision makers and decision subjects that result in tangible and intangible harms.
When harms happen, doctors, patients, insurers and others will turn to legal representatives to discover and argue for the extent to which accountability lies with different parties---the decision maker, decision subject, the entity that deployed the AI system (e.g., hospital system), or entity that developed the system.
Once a lawsuit is raised, there is a different stage of vulnerability where recourse now resides in the ability of the legal representative to discover the facts of what happened within the AI system that support their case. When an AI system is at the core of the harm, ascertaining the relevant factors for a persuasive case can become more complicated than normal,
requiring access to proprietary information about the algorithms, models, and data used to develop the AI system. In these cases, a power imbalance exists in favor of the party with access to the system.

Although legal experts are stakeholders in the design of AI systems~\cite{Mansi2024_LawyersContestability}, 
further research is needed in how to incorporate legal expertise into the design of AI systems and explanation systems.
The incorporation of legal considerations into the design of AI explanations has largely focused on theoretical impacts of the law on the design of explanations. 
Legal scholars~\cite{GerkeEtAl2020,schwammberger2023vision,fraser2022ai} have discussed legal imperatives for explanations, what explanations should enable, and the forms that explanations should take. Scholars have also advocated for auditing~\cite{Price2022, AbramoffEtAl2020,guha2023ai} an algorithm's inner workings, outputs, or data, and the advantages and legal imperatives of each. However, there are significant challenges in understanding how to design explanations for use in the legal system. For example, studies have demonstrated that traditional counterfactual explanations from AI systems are not intuitive to judges and are limited in their ability to support understanding in a legal case \cite{YacobyEtAl2022_HCXAI}. This can significantly impact how lawyers are able to represent physicians and patients in court.

\section{Design Opportunities for Legal Considerations}

To protect physicians and patients from harmful outcomes from decisions made with AI systems, it is imperative to understand how to incorporate legal implications into medical AI explanations. We highlight two initial opportunities to begin addressing this need. 

The first opportunity is around investigating the kinds of legal information that can support physicians. As mentioned before, as decision makers, physicians are placed in positions of power and legal responsibility, acting as a first defense against negative AI determinations about patients' medical needs. \textit{Legally informative} information explicitly describes laws, regulations, and legal rights that are relevant to those impacted by the AI system. For example, this includes information that doctors are fully liable for harms from decisions made with in-house (i.e. hospital-made) AI systems in which they serve as the final decision maker \cite{Sundholm2024}. On the other hand, \textit{legally actionable} information can be used for legal action once harm has occurred, but is not legally related in and of itself. For example, doctors' documentation is not legal information, but can be used in a malpractice suit. Helping doctors understand what information from the AI system can become legally actionable for themselves and others can be critical for contestability and the broader development of responsible AI systems.

The second opportunity relates to evaluations of new XAI systems. Evaluations are touch points with reality: they ideally reflect and (re-)direct AI system designs to account for the kinds of legal information and considerations that users actively have. 
Decision makers, decision subjects, and legal representatives have differing positionalities and vulnerabilities with respect to AI systems and the power dynamics that are reinforced by the presence of said systems. Creating ``Legally-Informed'' XAI, necessitates legally-informed evaluations---evaluations that can help us understand and adjust how AI systems support users' legal considerations. Legally-informed evaluations must take into account several important considerations highlighted in prior HCXAI Workshops, including the inherently multi-stakeholder nature of critical environments \cite{SchneiderEtAl2024_HCXAI}; the importance of understanding and designing around the actions each stakeholder group can take \cite{Mansi2023DoSomething}; and the variation in expectations from explanations specific to different groups \cite{SmallEtAl2023_HCXAI, Conati2024_HCXAI}.

\section{Co-Disciplinary Next Steps}

How does one identify legally informative and legally actionable information?
How does one design XAI systems to incorporate this information? 
As a practical first step, we advocate for co-disciplinary analyses---where researchers, developers, lawyers, and other stakeholders actively work together---to understand patterns across current litigation and reported harms. 

One place to start is to analyze 
litigation and harms databases which reflect what kinds of situations are litigated (or not). 
Many law schools are maintaining databases tracking current litigation, such as Georgetown's Health Litigation Tracker \cite{Georgetown_LitigationTracker} or George Washington University's Database of AI Litigation \cite{DAIL}. Lawsuits take time to develop and litigate, and may not reflect all of the potential risks to users (i.e., doctors and patients). For example, Epic, the American healthcare software company that develops the most widely used electronic health record system in the United States, deployed an AI-based sepsis prediction model that was highly criticized for its poor performance and ultimately retracted because of how much it disrupted patient care \cite{WongEtAl2021}. While a lawsuit did not cause the system to be retracted, it still had a significant and negative impact on how doctors cared for patients. Consequently, we also recommend that interdisciplinary teams analyze crowd-sourced databases that report these kinds of harms as well, such as the AI, Algorithmic and Automation Incidents and Controversies Database \cite{AIAAIC}). 

Coordinated, co-disciplinary analyses of these cases by researchers, developers, and lawyers can help reveal technical, social, and legal challenges that users are facing as the impacts of AI systems ripple across users' sociotechnical environments. 
Co-disciplinary analyses of current litigation and harms can build the necessary human capacity for cross-disciplinary communication to effectively address users' legal considerations. Prior work has highlighted the challenges of and need for cross-disciplinary communication to align algorithmic and legal pathways for actionability and contestability \cite{Mansi2024_LawyersContestability}. For example, there are fundamental terminology differences among communities, such as what information from an AI system constitutes an ``explanation'' \cite{Mansi2024_LawyersContestability}. 
Joint conversations around litigation and harms can expose these communication differences and build the capacity among people in different fields to create a shared understanding through conversation, increasing the ability to identify and address disciplinary blind-spots \cite{AgreChapter1998}. 
Through a shared understanding, people can then together better design legal, technical, and social infrastructures that truly serve users' information and action needs, expanding their actionability and contestability.

Coordinated, co-disciplinary analyses around current litigation and harms can also positively support communication around users' needs in several ways. 
Conversations around litigation can engage co-disciplinary groups in creating practical solutions to on-the-ground challenges. Analyzing existing litigation and harms can ensure both conversations focus on actionability and contestability---two of the major outcomes for users that can positively support their abilities to respond to deployed AI systems. Centering conversations on legal harms and repercussions impacting users can productively direct conversations about the actions that users, and others supporting them, can take to advocate for themselves, and the information they need to help them take these actions. 
This can reveal the information to include in explanations based on what is legally significant in lawsuits. 

Tying these kinds of information to the actions users need---validating information from the system, comparing alternative outcomes from AI decisions, communicating about AI decisions or outcomes (among others)---can be one way to build better evaluations, too. By stating assumptions about how different kinds of information support users' legal considerations, evaluations can then be developed to test these assumptions and improve AI systems in response. Consistently evaluating AI systems and explanations can help adapt explanatory content to the dynamic, legal landscape that shapes users' sociotechnical environment. For example, co-disciplinary teams can conduct qualitative analysis around how lawyers leverage explanatory information, providing helpful feedback on how they may  incorporate this information when litigating a harm or contestation.

\section{Conclusions}

Without accounting for legal considerations in people's sociotechnical environments, AI explanations risk tipping the balance of power away from those who are vulnerable and those advocating for the vulnerable. \textit{Legally-Informed} XAI is needed to ensure the actionability and contestability of AI systems. Towards this end, we describe the need to account for the legal considerations of three stakeholder groups decision makers, decision subjects, and legal representatives. As an example of the complexities that Legally-Informed XAI may need to address, we engage in an extended discussion of legal considerations via the context of medical AI systems. We offer design opportunities to begin understanding and shaping Legally-Informed XAI to ensure AI systems are actionable and contestable by people impacted by their determinations.

\begin{acks}
This material is based upon work supported by the National Science Foundation GRFP under Grant No. DGE-2039655. Any opinion, findings, and conclusions or recommendations expressed in this material are those of the authors(s) and do not necessarily reflect the views of the National Science Foundation.
\end{acks}

\bibliographystyle{ACM-Reference-Format}
\bibliography{references}


\end{document}